\documentclass[
    ,final            
  ]
  {aipproc}

\layoutstyle{8x11single}

\newcommand{\ga}{\lower.5ex\hbox{$\; \buildrel > \over \sim \;$}}
\newcommand{\la}{\lower.5ex\hbox{$\; \buildrel < \over \sim \;$}}

\begin{document}

\title{Finding (or not) New Gamma-ray Pulsars with GLAST}


\author{Scott M.~Ransom}{
  address={National Radio Astronomy Observatory, 520 Edgemont Rd., 
    Charlottesville, VA, 22901, USA}
}

\begin{abstract}
  Young energetic pulsars will likely be the largest class of Galactic
  sources observed by GLAST, with many hundreds detected.  Many will
  be unknown as radio pulsars, making pulsation detection dependent on
  radio and/or x-ray observations or on blind periodicity searches of
  the gamma-rays.  Estimates for the number of pulsars GLAST will
  detect in blind searches have ranged from tens to many hundreds.  I
  argue that the number will be near the low end of this range, partly
  due to observations being made in a scanning as opposed to a
  pointing mode.  This paper briefly reviews how blind pulsar searches
  will be conducted using GLAST, what limits these searches, and how
  the computations and statistics scale with various parameters.
\end{abstract}

\maketitle
\section{Introduction}

Pulsar emission mechanism(s) from radio to gamma-rays are poorly
understood after 40\,yrs of work.  Yet despite this fact, pulsars are
high-precision tools that probe a variety of topics in both
fundamental physics and astrophysics.  EGRET detected pulsed emission
from at least 6, and probably 7$-$9, young pulsars at energies
$\ga$100\,MeV \citep{tho04}.  In addition, several tens of the
(primarily Galactic) unidentified EGRET sources are likely pulsars
\citep[e.g.][]{mc00, mc03, kbm+03}.  With its much larger effective
area and improved angular resolution, GLAST will almost certainly
detect hundreds of pulsars.  The majority of those pulsars will remain
unidentified though (and useless as physics tools), unless pulsations
are detected through 1) ``folding'' of the events using timing
ephemerides for known radio pulsars; 2) searches of associated x-ray
sources to find radio-quiet Geminga-like pulsars \citep{hh92}, or very
faint radio pulsars \citep{hcg+01,rhr+02}; or 3) ``blind'' searches of
the gamma-ray events \citep{ckl+01}.  This paper discusses the latter,
and perhaps most difficult, option.

The known gamma-ray pulsars have similar characteristics \citep[for a
review, see][]{tho04}.  They are mostly non-variable sources with
fairly flat energy spectra in the gamma-ray regime, but with spectral
cutoffs around or just above 1\,GeV.  The flat energy spectra imply
that most of the {\em photons} detected by the Large Area Telescope
(LAT) will be in the $\sim$100-300\,MeV range, where the point spread
function has a width $\ga$1$^\circ$. The gamma-ray pulse shapes are
complex with two relatively sharp pulses which provide higher harmonic
content in searches and make their detection easier.

Most of the GLAST mission will be spent in sky survey mode, where the
whole sky is scanned every $\sim$3 hours\footnote{For details, see the
  GLAST mission website: \url{http://glast.gsfc.nasa.gov}}.  On
average, a point in the sky will be within the LAT field of view
$\sim1/6$ of the time.  Pointed observations would increase this
fraction to $\sim1/2$ (with Earth occultations preventing higher
on-source efficiency).  This loss in efficiency due to scanning (more
specifically, the resulting decrease in the number of source events
$N_s$ during a particular viewing period $T_{\rm view}$) will make
coherent pulsation searches considerably more difficult and less
sensitive.

\section{Blind Searches for Gamma-ray Pulsars}

Blind searches of low count rate event data are typically conducted
using either Fourier techniques (i.e.~binning events into a time
series, computing one or more FFTs, and analyzing the resulting
amplitude and possibly phase spectra; \citep[e.g.][]{van89,ckl+01}) or
via brute-force epoch folding (i.e.~assembling a pulse profile by
determining the pulse phase of each event from a trial ephemeris and
then computing a probability of non-uniformity for the profile;
\citep[e.g.][]{drs89,gl92}).  Optimal sensitivity to pulsations comes
from searches that treat all of the data in a ``coherent'' fashion,
meaning that the pulsar's rotational phase is accurately tracked over
the full observation or analysis duration $T_{\rm view}$, from first
event to last.

\begin{figure}
  \includegraphics[height=.34\textheight]{combined_fdot_effects.eps}
  \hspace{1cm}
  \includegraphics[height=.34\textheight]{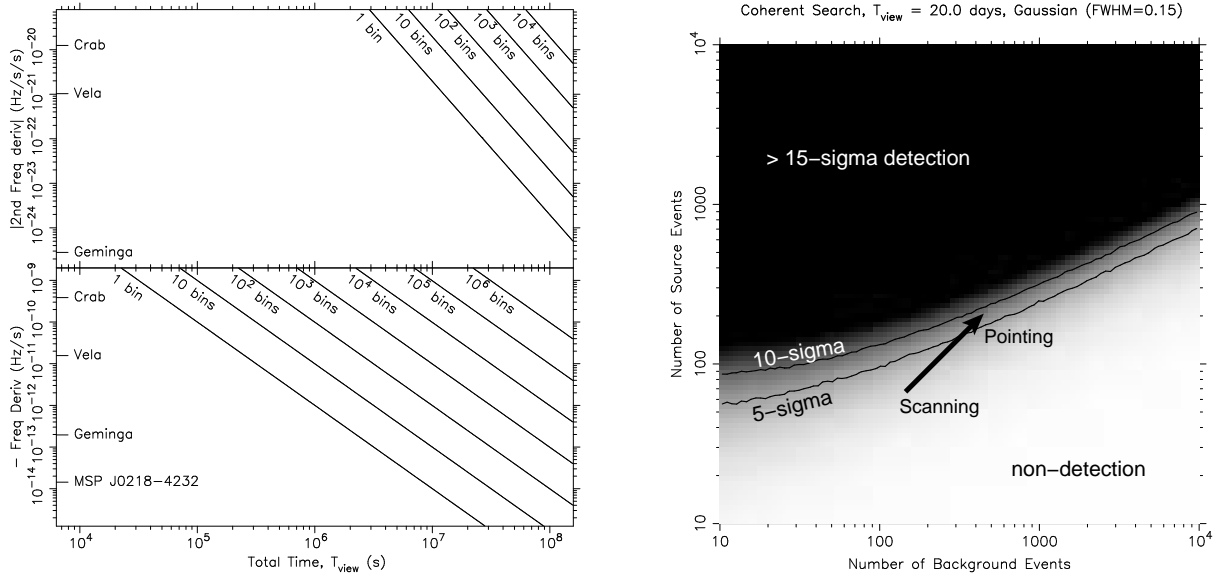}
  \caption{{\bf Left)} The number of independent Fourier ``bins''
    (i.e. 1/$T_{\rm view}$ in Hz) a signal with frequency derivative
    $\dot f$ (bottom) and frequency 2$^{\rm nd}$ derivative $\ddot f$
    (top) would drift in an observation of duration $T_{\rm view}$.
    To preserve sharp features in pulse profiles, search codes need to
    account for signal drift to a small fraction (1$-$10\%) of a
    Fourier bin.  {\bf Right)} The 95\% confidence-level detectability
    of a blind event-folding search with $T_{\rm view}$=20\,days for a
    pulsar with a Gaussian pulse profile with FWHM=15\% in pulse
    phase.  The ``sigmas'' represent Gaussian significance in standard
    deviations after accounting for the number of trials searched.
    The arrow (representing a factor of $\sim$3 increase in
    accumulated events during $T_{\rm view}$ and which can be
    translated anywhere on the plot) shows how the significance
    improves for pointed observations (which are still affected by
    Earth occultations) as opposed to scanning observations.  In this
    case, a 20-day pointing results in an $\sim$8-$\sigma$ detection,
    whereas 20-days of scanning gives a non-detection.}
\end{figure}

The real problem with gamma-ray pulsation searches comes from the fact
that the sources have very low count rates of
10$^{-7}$$-$10$^{-8}$\,photons($>$100\,MeV)\,cm$^{-2}$\,s$^{-1}$,
which for the LAT corresponds to roughly 0.2$-$2 events per hour.
Therefore, very long observations lasting from weeks to years are
required to make significant detections.  And unfortunately, young
pulsars are notoriously badly behaved over such timescales.

Ideally, we would only need to search over the unknown spin frequency
$f$ of a pulsar.  However, young pulsars rapidly spin-down (i.e.~have
large frequency derivatives $\dot f$), they exhibit timing noise which
manifests itself in searches as significant frequency 2$^{\rm nd}$
derivatives $\ddot f$ (and for the noisiest pulsars, many higher order
derivatives as well), and they occasionally, on month or year
timescales, ``glitch'' and instantaneously change their observed $f$
and $\dot f$.  Figure~1 shows the number of independent frequency bins
of width $1/T_{\rm view}$ in Hz, corresponding to an accumulated phase
error of one rotation, over which a pulsar signal will ``drift''
during a coherent search for a variety of realistic values of $f$ and
$\ddot f$.  Pulsar spin-down becomes important for energetic young
pulsars in a day or two, and so all gamma-ray searches must account
for an unknown $\dot f$.  Timing noise can become important after a
few weeks or months.  Since gamma-ray pulse profiles contain sharp
features, and therefore their pulsed signals contain many harmonics,
search codes must significantly oversample the $\sim1/T_{\rm view}$ or
$\sim1/T_{\rm view}^2$ spacings in $f$ and $\dot f$ to maintain
optimal sensitivity.

A potentially bigger problem, as shown by \citet{ckl+01}, are
positional errors or uncertainties $\epsilon$, which cause {\em
  apparent} changes in the observed spin frequency $\delta\!f \sim
10^{-6}\epsilon_{\rm mrad}f_{10}\sin\theta$\,Hz, frequency derivative
$\delta\!\dot f \sim 2\times10^{-13}\epsilon_{\rm
  mrad}f_{10}\cos\theta$\,Hz\,s$^{-1}$, and frequency 2$^{\rm nd}$
derivative $\delta\!\ddot f \sim 4\times10^{-20}\epsilon_{\rm
  mrad}f_{10}\sin\theta$\,Hz\,s$^{-2}$, where $\epsilon_{\rm mrad}$ is
the position error in milliradians, $f_{10}$ is the spin frequency in
tens of Hz, and $\theta$ is the time varying angle between the Earth's
velocity vector and the line of sight to the pulsar.  For candidate
GLAST pulsar sources, $\epsilon$ will be milliradians (arcminutes) in
size, making the blind detection of even stable Geminga-like pulsars
much more difficult.

{\em The intrinsic $\dot f$, timing noise, glitch frequency, and
  position error effects place upper limits on the useful $T_{\rm
    view}$ for coherent pulsation searches to durations as short as
  $\sim$10\,days for the most energetic and active pulsars, or a few
  months for older, slower, and more stable pulsars.}  Because of the
scanning nature of the GLAST mission, there will be fewer counts
accumulated for a particular source over the specified $T_{\rm view}$
by a factor of $\sim$3 as compared to a pointed observation.  This
decrease in accumulated counts per unit time greatly impacts coherent
search sensitivity and is the reason why scanning is not the optimal
observing mode for pulsar studies.

\subsection{Pulsation Detection and Computational Considerations}

The power $P$ in a periodic gamma-ray signal is $P \sim 1 + \alpha
N_s^2/N_t$, where $\alpha$ is a pulse-shape dependent factor
(0.4$-$0.9 for the known gamma-ray pulsars)\citep[e.g.][]{bos87}. The
total number of events $N_t$ is the sum of the source (pulsed) events
$N_s$ and background (unpulsed) events $N_b$.  The {\em probability}
that noise fluctuations produce a certain search power $P_o$ is {\em
  exponentially} related to the power: ${\rm Prob}(P\ge
P_o)\propto\exp(-P_o)N_{\rm trials}$, where $N_{\rm trials}$ is the
number of independent trials searched.  $N_{\rm trials}$ is
proportional to the range of $f$ and $\dot f$ searched, but more
importantly, to $T_{\rm view}^3$.

In general then, to optimize search sensitivity, we want large $N_s$
to maximize $P$, yet a relatively short $T_{\rm view}$ (appropriate
for the limits discussed in the last section) to reduce $N_{\rm
  trials}$ and the computational complexity.  This rule exposes
another problem for scanning-mode observations of pulsars: if $T_{\rm
  view}$ is increased by a factor of $\sim$3 to recover the events
``lost'' compared to a pointed observation (assuming the pulsations
remain well-behaved over the longer $T_{\rm view}$), a detected signal
will have the same power as from the shorter pointed observation, yet
the probability that the signal is significant will decrease by a
factor of $\sim3^3 = 27$ due to the larger $N_{\rm trials}$ searched.

Computationally, the complexity of epoch-folding searches scales
roughly as $T_{\rm view}^4$ (i.e.~$N_t\times N_{\rm trials}$), while
FFT-based techniques fare slightly better and scale as $T_{\rm
  view}^3\log T_{\rm view}$.  Moderate sized computer clusters can
currently handle epoch-folding searches of hundreds to thousands of
events for viewing periods of up to several weeks (see Figure~1).  For
longer $T_{\rm view}$ or more limited computing resources, one can use
evolutionary \cite{bk96} and/or heirarchical \cite{ran02} techniques
to greatly reduce the computational burden of coherent searches
without greatly impacting the overall sensitivity, {\em if} the
average pulsation flux is time invariant.  Coherent searches with
GLAST with $T_{\rm view}\la1$$-$2\,months should be sensitive to
pulsars with photon fluxes ($>$100\,MeV) of a few times 10$^{-8}$ to a
few times 10$^{-7}$\,photons\,cm$^{-2}$\,s$^{-1}$ for scanning-mode
observations, or as low as
$\sim$2$\times$10$^{-8}$\,photons\,cm$^{-2}$\,s$^{-1}$ for pointed
observations.

Finally, incoherent searches are possible where the absolute phase of
pulsations is {\em not} tracked over the full $T_{\rm view}$, but only
over much shorter ``windows'' $T_{\rm win}$.  Incoherent searches are
less sensitive than coherent ones for the same $T_{\rm view}$, but
they use many fewer computing resources and enable searches with much
longer $T_{\rm view}$ \cite{van89,azjb06}.  The basic idea limits the
duration of the coherently processed intervals $T_{\rm win}$ such that
a signal with any reasonable $\dot f$ or $\ddot f$ stays within a
single independent frequency bin (of width $1/T_{\rm win}$\,Hz) or
fraction thereof.  These windowed analyses are then combined without
phase information, but with corrections for $\dot f$ and/or $\ddot f$
effects occurring between the intervals, and analyzed.  Initial
simulations show that one might detect {\em relatively stable pulsars}
with photon fluxes ($>$100\,MeV) as low as
1$-$2$\times$10$^{-8}$\,photons\,cm$^{-2}$\,s$^{-1}$ with $T_{\rm
  view}\sim1$\,yr and $T_{\rm win}\sim1$\,day.  Without phase
information, however, the folding of all events over such long
durations to accumulate high-quality pulse profiles may be difficult
or even impossible.

In summary, instabilities of young pulsars, source positional
uncertainties, and the default survey/scanning mode for the mission
will make blind pulsation searches with GLAST much more difficult than
early estimates had predicted.  It is therefore likely that GLAST will
blindly discover only tens of new pulsars rather than hundreds.

{\bf Acknowledgements:} Thanks go to Paul Ray, Julie McEnery, Mallory
Roberts, Maura McLaughlin, and Kent Wood for useful discussions.


\end{document}